\documentstyle[osa,multicol,aps,graphicx,epsfig]{revtex}

\begin{document}
\title{Stable oscillating nonlinear beams in square-wave-biased-photorefractives}
\author{Giorgio Maria Tosi-Beleffi, Marco Presi, Claudio Palma, Danilo Boschi and Eugenio DelRe}
\address{Fondazione Ugo Bordoni, Via B. Castiglione 59, 00142 Rome, Italy}
\address{Istituto Nazionale Fisica della Materia, Unita' di Roma 1, 00100 Rome, Italy}
\address{Dipartimento di Fisica, Universita' Roma Tre, 00146 Rome, Italy}
\author{Aharon J. Agranat}
\address{Applied Physics Department, Hebrew University of Jerusalem, Jerusalem 91904, Israel}
\date{\today}
\maketitle

\begin{abstract}
\noindent We demonstrate experimentally that in a centrosymmetric
paraelectric non-stationary boundary conditions can dynamically
halt the intrinsic instability of quasi-steady-state
photorefractive self-trapping, driving beam evolution into a
stable oscillating two-soliton-state configuration.
\end{abstract}


\begin{multicols}{2}
\narrowtext The propagation of light in a photorefractive crystal
gives rise to intense beam self-action that, in its most generic
manifestation, causes fanning and anisotropic scattering
\cite{general1}. The application of an external bias field to the
crystal can drastically change this behaviour, and allow spatial
self-trapping and soliton formation: the non-diffracting
propagation of micron-sized optical beams \cite{general2}. In a
biased system this occurs only in a transient regime, for a finite
time window, and the resulting nonlinear waves are called
quasi-steady-state-solitons \cite{quasi-steady-state}.  By
increasing the natural dark conductivity, this regime can be made
stable, giving rise to steady-state screening solitons
\cite{screening}.

In this paper we investigate, for the first time, a fundamentally
different stabilization process connected to beam behaviour in a
non-stationary external bias field \cite{nonstationary}. In
particular, we study beam evolution in the presence of an
alternating field in centrosymmetric
potassium-lithium-tantalate-niobate (KLTN) \cite{KLTN}, a material
known to support a rich variety of nonlinear beam phenomena
\cite{euslab} \cite{euneedle} \cite{eudiffusion} \cite{euspont}.

Results indicate that, for appropriate conditions, the beam
self-trapping process, that leads to transient quasi-steady-state
solitons for stationary conditions, can be driven into a stable
self-trapped regime, formed by an alternating oscillation between
two beam trajectories, in the {\bf{\it absence}} of enhanced dark
conductivity. This phenomenon, in our understanding, is made
possible by the fact that single optical trajectories,
corresponding to the two alternate states of the bias field,
non-coincident due to asymmetric diffusion-seeded bending and
electro-optic read-out effects, engender the simultaneous
formation, through the quadratic electro-optic response of the
crystal in the paraelectric phase, of two trapping index patterns,
that form two back-to-back specular double layers of charge that
halt runaways charge buildup.

Experiments are carried out in samples of zero-cut centrosymmetric
photorefractive KLTN, a composite perovskite doped with Copper and
Vanadium impurities, with a set-up that is similar to those
generally used in photorefractive soliton studies \cite{euslab}
 \cite{euneedle}, apart from the absence of background illumination and
the use of an alternating external voltage source.  The sample
temperature is kept at a given value T by means of a stabilized
current controlled Peltier-junction. A $\lambda$=514nm
continuous-wave TEM$_{00}$ beam, from an argon-ion laser, is first
expanded and then focused onto the input facet of the sample, and
launched along the crystal principal axis z. Focusing is obtained
either with a cylindrical y-oriented lens, giving rise to a
one-dimensional beam confined in the x transverse direction, for
investigation of slab-solitons, or a spherical lens for full
two-transverse-dimensional (i.e. x and y) investigation of
needle-solitons. The electrodes are deposited on the x facets and
the source can provide a square-voltage wave-form of variable
peak-to-peak amplitude V$_{sq}$ and period T$_{sq}$. Beam dynamics
are observed by means of a top-view and a transverse CCD camera.

\begin{figure}
\begin{center}
\resizebox{8.5cm}{!}{\mbox{\includegraphics*[1cm,1cm][29cm,20cm]{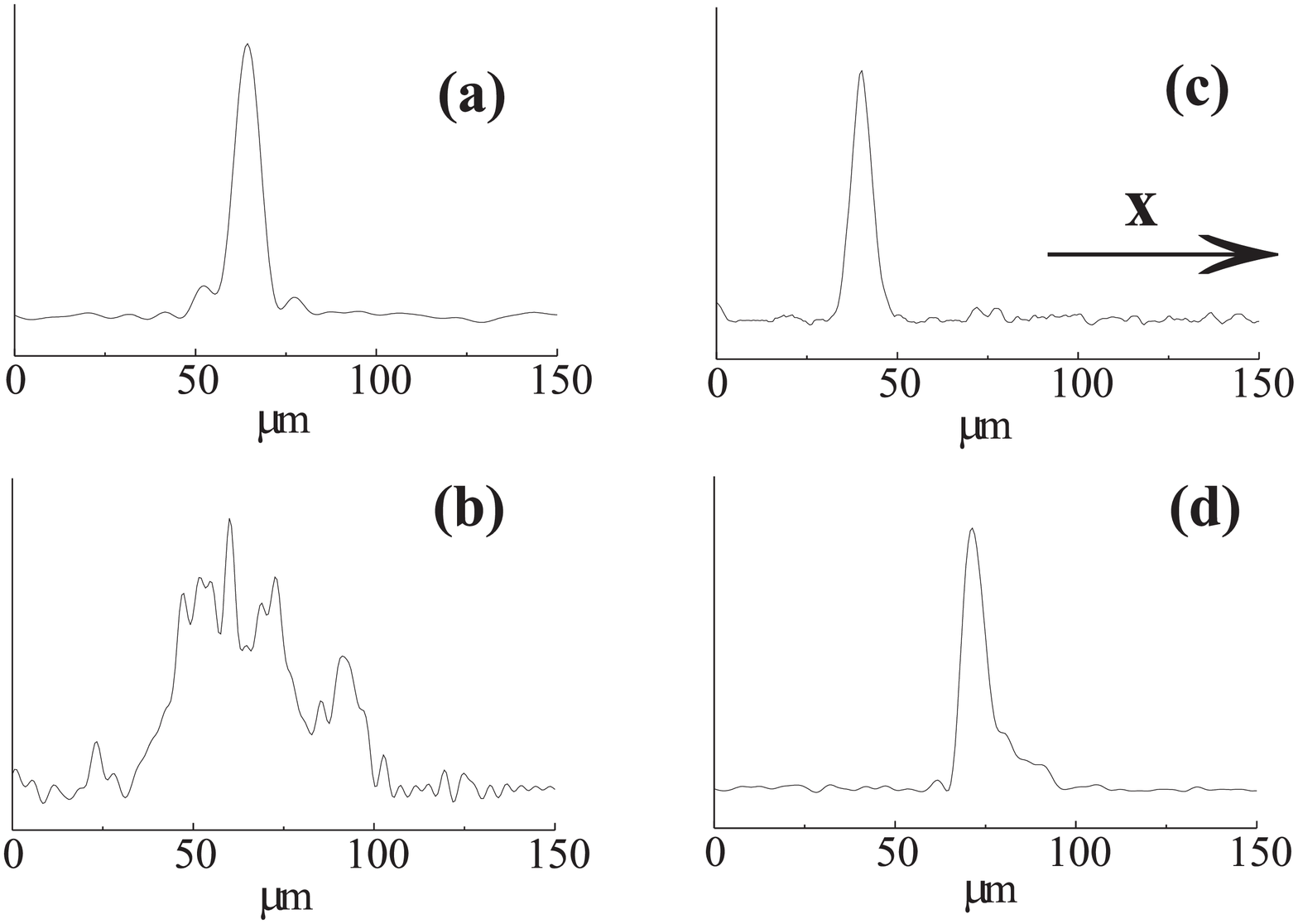}}}
\caption{Stable two-state oscillation of a slab-soliton beam
subject to a square-wave bias. (a) Normalized input intensity
profile; (b) Output intensity profile after 2.4mm linear
propagation in the crystal; (c-d)  Output intensity self-trapped
profiles of the two alternating states.}
\label{oscillation-profile-1D}
\end{center}
\end{figure}

The main qualitative phenomenology observed is contained in
Fig.(\ref{oscillation-profile-1D}), where the two-branch
oscillation along the x-axis is shown. The laser beam is focused
by means of an f=150mm cylindrical lens onto the input facet of a
3.7$^{(x)}$x4.7$^{(y)}$x2.4$^{(z)}$ mm sample, giving rise to an
approximately one-dimensional (transverse dimension x) fundamental
Gaussian beam diffracting in the x direction, as the beam evolves
along z. The input beam has a full-width-half-maximum (FWHM) of
$\simeq$ 5 $\mu$m, and diffracts to 44 $\mu$m after propagating
2.4 mm in the sample (n$\simeq$ 2.4) (see
Fig.(\ref{oscillation-profile-1D}(a)-(b)). The beam has a peak
intensity I$_{p}$ $\simeq$ 3kW/m$^2$, whereas no background
illumination is implemented.  In these conditions, but with a
stationary bias, quasi-steady-state self-trapping is observed
after a response time $\tau_1 \simeq$ 3min, for an external
voltage on the x electrodes of V=380V.  In the oscillating
configuration of Fig.(\ref{oscillation-profile-1D}(c)-(d)), the
external square-wave bias voltage has a peak to peak amplitude
V$_{sq}$=760V with a period T$_{sq}$= 10 s (i.e., T$_{sq} \ll
\tau_1$), the crystal being kept at T=18$^{\circ}$C (thus having a
relative dielectric constant of $\epsilon_r \approx 11 \cdot
10^3$). As the external oscillation occurs, the beam undergoes
spatial confinement in one direction towards one electrode
(Fig.(\ref{oscillation-profile-1D}c)), and then in the other,
towards the opposite electrode (along the transverse x direction)
(Fig.(\ref{oscillation-profile-1D}d)), oscillating between two
distinct optical trajectories. The remarkable result is that this
stable oscillating regime, which has a buildup transient $\tau_2
\simeq 15$ min considerably longer than $\tau_1$, continues {\it
indefinetly}, and is thus a stable oscillation in which the beam
is {\it continuously} confined, but in {\it two} distinct
alternating trajectories.  The switching time $\tau _{sw} \ll$
T$_{sq}$ is associated with the crystal charging time.  Apart from
this very short interval $\tau_{sw}$, not of photorefractive
nature, the beam is continuously {\it trapped}. The distance
between the two trapped beams at the output is $\Delta$x$\simeq$
30 $\mu$m.

\begin{figure}
\begin{center}
\resizebox{8.5cm}{!}{\mbox{\includegraphics*[14cm,10cm][16cm,11cm]{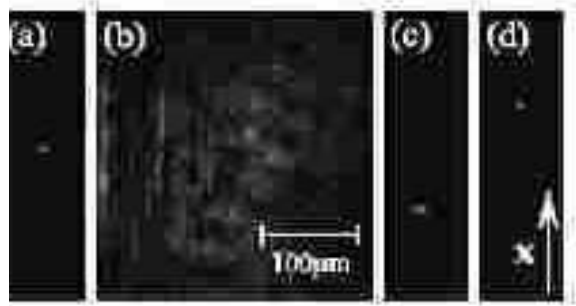}}}
\caption{Stable two-state oscillation of a needle-soliton beam
subject to a square-wave of peak to peak amplitude V$_{sq}$=700V,
of period T$_{sq}$=10s. (a) Input intensity 6 $\mu$m FWHM
distribution; (b) Output intensity 150 $\mu$m distribution after
6.4mm linear propagation in the crystal; (c-d)  Output intensity
self-trapped profiles of the two alternating states.}
\label{oscillation-image-2D}
\end{center}
\end{figure}

An analogous phenomenology is observed for two-dimensional
diffracting beams, and is shown in
Fig.(\ref{oscillation-image-2D}), in a second
2.2$^{(x)}$x2.2$^{(y)}$x6.4$^{(z)}$ sample of KLTN, kept at
T=26$^{\circ}$C ($\epsilon_r \approx$ 6.5$\cdot$10$^{3}$). Note
that the needle confinement extends over approximately 25
diffraction lengths.

As opposed to screening soliton phenomena, we found no strict
existence condition associated with the electro-optic response
(i.e., V$_{sq}$ and crystal T), much like standard
quasi-steady-state self-trapping experiments
\cite{quasi-steady-state}. The only observable difference in the
final oscillating state that depends appreciably on changes in the
electro-optic response is the divergence angle of the two
trajectories, that gives rise, at the output, to a different value
of $\Delta$x.  We found that a stronger static polarization
induces a stronger divergence.

\begin{figure}
\begin{center}
\resizebox{8.5cm}{!}{\mbox{\includegraphics*[13cm,9cm][17cm,11.5cm]{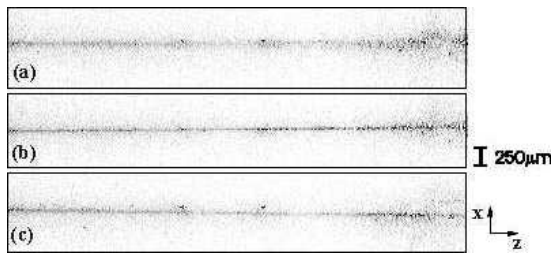}}}
\caption{Soliton swinging: Slab-beam transient dynamics for a
$\tau_1 \simeq$2s  and T$_{sq}$=10s in a 6.4mm long sample of
KLTN, with a V$_{sq}$, sample T=27$^{\circ}$C.  (a) Top view (y
direction) of the diffracting beam, with a 6$\mu$m input FWHM, and
linear diffracting 120$\mu$m output; (b)-(c) Two opposite
transient self-trapped states.} \label{top-view-1D}
\end{center}
\end{figure}

On the other hand, we observed a distinct dependence of beam
evolution on the time scales involved, suggesting that the main
underlying mechanism is strongly connected to the temporal
oscillations in the boundary conditions. We thus carried out
experiments changing the nonlinear time constant, connected to
beam peak intensity I$_p$, keeping all other parameters unaltered.
Thus, we essentially vary the ratio T$_{sq}$/$\tau_1$, since
$\tau_1$ is approximately proportional to I$_p$.  For slow enough
dynamics i.e., for T$_{sq}$/$\tau_1 \ll$ 1,  the steady
oscillating state is {\it always} reached (at least for the
investigated cases), whereas for very rapid dynamics, i.e., for
beam intensities such that T$_{sq}$/$\tau_1 \gg$ 1 , single branch
evolution is allowed to reach quasi-steady-state destabilization
and the steady oscillating situation is not observed. In this
case, the beam undergoes a distinctive swinging evolution
mimicking the two-state self-trapping of the previous case, as
shown in Fig.(\ref{top-view-1D}). During one field oscillation,
the beam first undergoes self-trapping, deflected in one direction
(Fig.(\ref{top-view-1D}b)), decays (see Fig.(\ref{top-view-1D}a)),
diffracting in the forward z direction, and then forms a second
transient soliton in the opposite direction
(Fig.(\ref{top-view-1D}c)), decays again in the z direction, and
so forth.  Thus, in this case, no {\it stable} self-trapped
oscillating configuration is reached, since most of the time the
beam is diffracting (T$_{sq} \gg \tau_1$). Note that the swinging
motion is a direct consequence of the residual charge displacement
of the previous state, and is much more pronounced than single
beam self-bending observed in stationary conditions
($\Delta$x$\simeq$30$\mu$m as compared to $\Delta$x$\simeq$5$\mu$m
of conventional self-bending observed).

\noindent We shall limit our discussion to slab-solitons, since
understanding of even basic needle soliton phenomenology is still
unclear \cite{unclear}. Concerning the formation of the two-state
oscillation, we note that, although in a purely drift-like
configuration free photoexcited charge under the influence of a
zero average square-wave alternate field E$_{sq}$, with $T_{sq}\ll
\tau _{1}$, cannot give rise to any space-charge separation,
asymmetric charge diffusion components can seed two intensity
distributions I$_{+}$ and I$_{-}$, corresponding to the two
alternate states of external field, that separate during
propagation along the z axis, and thus allows a non-zero
photorefractive response \cite{self-bending}.  The final state is
a product of beam charge separation during one electric field
polarity, combined with electro-holographic effects during the
opposite polarity phase \cite{eu}.

\begin{figure}
\begin{center}
\resizebox{9cm}{!}{\mbox{\includegraphics*[13cm,10cm][17cm,11cm]{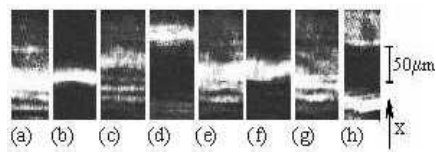}}}
\caption{Break-down of the two-state formation process. Output
intensity distribution (conditions of
Fig.(\ref{oscillation-profile-1D})). (a) Initial output
diffraction for V=0; (b) Quasi-steady-state self-trapping for
V=V$_{+}$; (c) Read-out intensity for V=0 (after soliton
formation); (d) Read-out intensity for V=V$_{-}$; (e,f,g,h)
specular results starting from V=V$_{-}$.}
\label{process-breakdown}
\end{center}
\end{figure}

\noindent To break down the process, consider the formation of a
single quasi-steady-state soliton with a constant voltage
V=V$_{+}$ (equal to the positive value of the alternating field in
the oscillating case)(see Fig.(\ref{process-breakdown}a-b)). If we
halt beam evolution (attenuating the propagating beam intensity)
before the transient trapped regime has decayed (i.e., for
t$<\tau_{1}$), and put V=0, we observe an increased beam
diffraction at the output, a signature of the residual defocusing
pattern (Fig.(\ref{process-breakdown}c)).  The diffraction is
furthermore slightly asymmetric, as a consequence of the diffusion
component in charge separation. If we invert the applied electric
field (i.e. V=V$_{-} \equiv $-V$_{+}$), the defocusing pattern is
enhanced, and most of the light is diverted to a limited region
strongly shifted with respect to the input beam
(Fig.(\ref{process-breakdown}d)).  A wholly specular behaviour is
observed when the quasi-steady-state soliton is originally formed
with a constant V=V$_{-}$ (Fig.(\ref{process-breakdown}e,f,g, and
h)). In the actual alternating field case, since we start from a
zero charge separation, and do not allow the system to evolve, if
not only partially, during each oscillation ($T_{sq}\ll \tau
_{1}$), the final charge separation and index pattern will be a
{\it symmetric} hybrid combination of the two self-trapped and two
deflected beams (see Fig.(\ref{process-breakdown}b,d,f,and h)),
whereas the actual beam trajectory will switch, following the
oscillation of the external field.  Note that, during the entire
oscillation, the beam continuously maintains its confinement,
apart from the small transient $\tau_{sw}$.

Concerning the stability of the nondiffracting pattern, in absence
of artificial dark illumination we note that in our case charge
separation is intrinsically symmetric. Mobile electrons are forced
to drift towards the central region between the two trajectories,
forming a potential barrier to further charge separation and
saturation, effectively freezing the space-charge structure. The
switching, in this case, is only of electro-optic nature, and
occurs as a consequence of the electro-holographic read-out of the
asymmetric index components in the initial stages of propagation
(where the two trajectories are superposed), after the
space-charge field has reached a steady-state.

The work of E.D. and M.P. was carried out in the framework of an
agreement between Fondazione Ugo Bordoni and the Italian
Communications Administration. Research carried out by A.J.A. is
supported by a grant from the Ministry of Science  of the State of
Israel. We thank Luigi Piccari for useful discussions.

\end{multicols}


\begin{thebibliography} {aa}

\bibitem{general1} L. Solymar, D. J. Webb, and A. Grunnet-Jepsen,
{\it The physics and applications of photorefractive materials}
(Clarendon Press, Oxford 1996).

\bibitem{general2} M. Segev and M. Stegeman, Phys.
Today {\bf 51}, 42 (1998); G.I. Stegeman and M. Segev, Science
{\bf 286}, 1518 (1999).

\bibitem{quasi-steady-state} G.C. Duree, J.L. Shultz, G.J. Salamo,
M. Segev, A. Yariv, B. Crosignani, P. Di Porto, E.J. Sharp, and
R.R. Neurgaonkar, Phys.Rev.Lett. {\bf 71}, 533 (1993).

\bibitem{screening} M. Segev, G. Valley, B. Crosignani, P. Di
Porto, and A. Yariv, Phys.Rev.Lett. {\bf 73}, 3211 (1994); D.N.
Christodoulides and M.I. Carvalho, J.Opt.Soc.Am.B {\bf 12}, 1628
(1995).

\bibitem{nonstationary}In linear
configurations see, e.g., A. Grunnet-Jepsen, C.H. Kwak, and L.
Solymar, Opt.Lett. {\bf 19}, 1299 (1994).

\bibitem{KLTN} A.J. Agranat, R. Hofmeister, and
A. Yariv, Opt. Lett. {\bf 17}, 713 (1992).

\bibitem{euslab}  E. DelRe, B. Crosignani, M. Tamburrini, M.
Segev, M. Mitchell, E. Refaeli, and A.J. Agranat, Opt.Lett. {\bf
23}, 421 (1998).

\bibitem{euneedle} E. DelRe, M. Tamburrini, M. Segev, E. Refaeli,
and A.J. Agranat, Appl.Phys.Lett. {\bf 73}, 16 (1998).

\bibitem{eudiffusion} B. Crosignani, A. Degasperis, E. DelRe, P.
Di Porto, and A.J. Agranat, Phys.Rev.Lett. {\bf 82}, 1664 (1999).

\bibitem{euspont} E. DelRe, M. Tamburrini, M. Segev, R. Della
Pergola, and A.J. Agranat, Phys.Rev.Lett. {\bf 83}, 1954 (1999).

\bibitem{unclear} S. Gatz and J. Herrmann,
Opt.Lett. {\bf 23}, 1176 (1998); M. Saffman, A.A. Zozulya,
Opt.Lett. {\bf 23}, 1579 (1998).

\bibitem{self-bending} M. Carvalho, S. Singh, D. Christodoulides ,
Opt.Comm. {\bf 120}, 311 (1995).

\bibitem{eu} E. DelRe, M. Tamburrini and Aharon J. Agranat,
Opt.Lett. {\bf 25}, 963 (2000)



\end{thebibliography}
\end{document}